\documentclass[aps,prl,twocolumn,superscriptaddress,showpacs,showkeys]{revtex4-1}
\usepackage{amsmath,latexsym,amsfonts,amssymb}
\usepackage{natbib}
\usepackage{epsfig}
\usepackage{graphicx}
\usepackage{inputenc}
\usepackage{color}
\usepackage{setspace}
\usepackage{multirow}
\usepackage{array}
\usepackage{tabularx}
\usepackage{bm}
\usepackage{color}
\usepackage{xcolor}
\usepackage{gensymb}
\usepackage{xcolor}
\usepackage[left]{lineno}
\usepackage{float} 
\newcommand{\bfs}{BaFe$_2$Se$_3$}
\newcommand{\bfsd}{BaFe$_2$S$_3$}

\newcommand{\jun}{$J_{1}$}
\newcommand{\Jpun}{$J^{'}_{1}$}
\newcommand{\Jdeux}{$J_{2}$}
\newcommand{\Jpdeux}{$J^{'}_{2}$}
\newcommand{\Jtrois}{$J_{3}$}
\newcommand{\Jcinq}{$J_{5}$}
\newcommand{\Jquatre}{$J_{4}$ }

\begin{document}

\title{Purely antiferromagnetic frustrated Heisenberg model in spin ladder compound BaFe$_2$Se$_3$}

\author{A. Roll}
\affiliation{Universit\'e Paris-Saclay, CNRS, Laboratoire de Physique des Solides, 91405, Orsay, France.}
\affiliation{Laboratoire L\'eon Brillouin, CEA, CNRS, Universit\'e Paris-Saclay, 91191, Gif sur Yvette, France}

\author {S. Petit}
\affiliation{Laboratoire L\'eon Brillouin, CEA, CNRS, Universit\'e Paris-Saclay, 91191, Gif sur Yvette, France}

\author{A. Forget}
\affiliation{Universit\'e Paris-Saclay, CEA, CNRS, SPEC, 91191, Gif-sur-Yvette, France.}

\author{D. Colson}
\affiliation{Universit\'e Paris-Saclay, CEA, CNRS, SPEC, 91191, Gif-sur-Yvette, France.}

\author{A. Banerjee}
\affiliation{Neutron Scattering Division, Oak Ridge National Laboratory, Oak Ridge, Tennessee 37831, USA}
\affiliation{Department of Physics and Astronomy, Purdue University, West Lafayette, IN 47906, USA}

\author{P. Foury-Leleykian}
\affiliation{Universit\'e Paris-Saclay, CNRS, Laboratoire de Physique des Solides, 91405, Orsay, France.}

\author{V. Bal\'edent}
\affiliation{Universit\'e Paris-Saclay, CNRS, Laboratoire de Physique des Solides, 91405, Orsay, France.}
\email [Corresponding author: ] {victor.baledent@universite-paris-saclay.fr}

\date{\today}

\begin{abstract}
The spin dynamics in the block magnetic phase of the iron-based ladder compound \bfs\ has been studied by means of single crystal inelastic neutron scattering. Using linear spin wave theory and Monte-Carlo simulations, our analysis points to a magnetic Heisenberg model with effective frustrated antiferromagnetic couplings only, able to describe both the exotic block order and its dynamics. This new and purely antiferromagnetic picture offers a fruitful perspective to describe multiferroic properties but also understand the origin of the stripe-like magnetic instability observed under pressure as well as in other parent compounds with similar crystalline structure.
\end{abstract}

\maketitle

\begin{figure}[!t]
    \includegraphics[width=1\linewidth, angle=0]{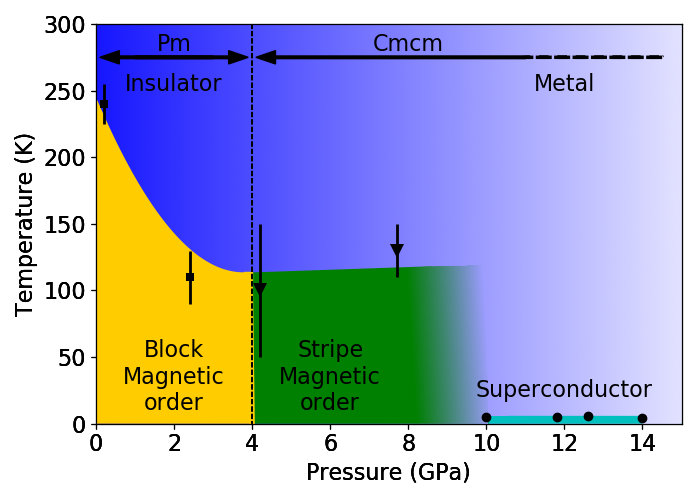}
    \caption{(Color online) Schematic pressure temperature phase diagram of \bfs. The block order (up-up-down-down) occupies the yellow region (black squares), while the stripe order (up-down-up-down) is observed in the green region (black triangles). Superconductivity appears above 10 GPa (black cricles from refs. \cite{Ying2017, zheng2022universal}). The structural transition from Pm to Cmcm space group is observed at 4 GPa, and progressive metallization occurs between 4 and 10 GPa.}
    \label{diagram}
\end{figure}

Magnetism plays a prominent role in many exotic properties of condensed matter, from unconventional superconductivity to multiferroicity and spin liquid phases. Copper has long been the focus of attention for studying these exotic phases, thanks to its half-integer spin, which promotes the emergence of these quantum states. On the other hand, research on multiferroicity has focused on manganese-based compounds, whose large magnetic moment has made it possible to obtain sufficiently strong magnetoelectric couplings to allow applications. More recently, iron-based compounds have found a place at the interface between these two paradigms. Indeed, unconventional superconductivity has been discovered in pnictides, and the large iron magnetic moment combined with magnetic interactions allows to obtain multiferroic materials at quite high temperatures. \bfs\ embodies this new path, by showing multiferroic properties below 250K, and superconductivity above 10 GPa (see phase diagram Fig.~\ref{diagram}). 

At ambient pressure, \bfs\ crystallizes in the polar Pm space group \cite{Zheng2020,Weseloh2022}, with a weak distorsion from the average Pnma space goup. For sake of clarity, we shall use the orthorhombic average space group, in which there are two ladders per unit cell, each formed by two adjacent legs running along the $b$ axis. Below $T_N=$ 250~K, an exotic magnetic order develops, called "block order", consisting in an antiferromagnetic (AFM) arrangement of square blocks of four parallel iron spins, pointing mainly along the $a$ direction, i.e. perpendicular to the ladders. Along the legs, the structure thus shows an up-up-down-down pattern \cite{Caron2012}. This model has been refined using single crystal neutron diffraction, revealing a tilt of the moments with respect to the $a$ axis, and resulting in an umbrella-like magnetic order \cite{Zheng2022}. Under pressure, this exotic magnetic state gives way to a more conventional order consisting in stripes of an up-down-up-down pattern along the legs \cite{zheng2022universal}, and analogous to the one observed in parent compounds with similar crystalline structure. It is worth noting that this classical up-down-up-down pattern is also stable close to the superconducting dome (see dark-green region in Fig.~\ref{diagram}). 

Understanding the magnetic couplings, 
at the heart of the different properties of this compound, requires the study of its excitations. On the basis of a first inelastic neutron scattering measurement performed on powder samples, the authors of Ref.\cite{Mourigal2015} have suggested an alternation of ferromagnetic and antiferromagnetic couplings along the legs. Although this model clearly stabilizes the correct magnetic structure due to the absence of magnetic frustration, it raises several questions. While the exchange paths are admittedly different between up-up and up-down iron bonds along a given leg \cite{Weseloh2022}, they hardly justify the change in both sign and amplitude, hence the nature of the exchange interactions, as also underlined in theoretical literature on spin dynamics \cite{Herbrych2018}. Furthermore, the ferromagnetic nature of some interactions seems incompatible with the Goodenough-Kanamori orbital rules : as the Fe-Se-Fe angle (denoted by $\psi$) strongly departs from 90 degrees (see Tab. \ref{tab:expph}), the antiferromagnetic super-exchange is expected to prevail over the ferromagnetic direct coupling. Another physical argument in favour of purely antiferromagnetic bonds along the ladder is the vicinity of the classical antiferromagnetic up-down-up-down order observed under moderate pressure (4 GPa) and in all other members of the family presenting a similar crystalline structure, such as KFe$_2$Se$_3$, CsFe$_2$Se$_3$ and {\bfsd} \cite{Caron2012,hawai2015,Takahashi2015}. 

\begin{table}[!hb]
    \centering
    \begin{tabular}{cccccc}
    \hline 
      $\psi_{Fe-Se-Fe}$ & 66(7)  & 71(6)  & 65.8(19) & 101(3) & 105(3)\\
    \hline 
    Involved coupling & J$_{1}$ & J$^{'}_1$ & J$_R$ & J$_2$ & J$^{'}_{2}$\\ 
     \hline
   \hline 
    \end{tabular}
    \caption{Angles (in degrees) between the Fe-Se-Fe bonds for the different possible paths within the ladder, according to the structure published in ref. \cite{Zheng2020}. The couplings correspond to the exchange interactions affected by these angles, as represented in Fig. \ref{INS map}g.}
    \label{tab:expph}
\end{table}

In this letter we present new inelastic neutron scattering (INS) measurements performed on a single crystal of \bfs\ providing unique details on the spin wave dispersion. The modeling of the data leads to a purely antiferromagnetic Heisenberg Hamiltonian, allowing one to describe both the magnetic order and the dynamics. This picture also provides a natural understanding of the evolution of the magnetism under pressure as well as the magnetic ground state of parent compounds.

\begin{figure*}[!th]
    \centering
    \includegraphics[width=1
    \linewidth, angle=0]{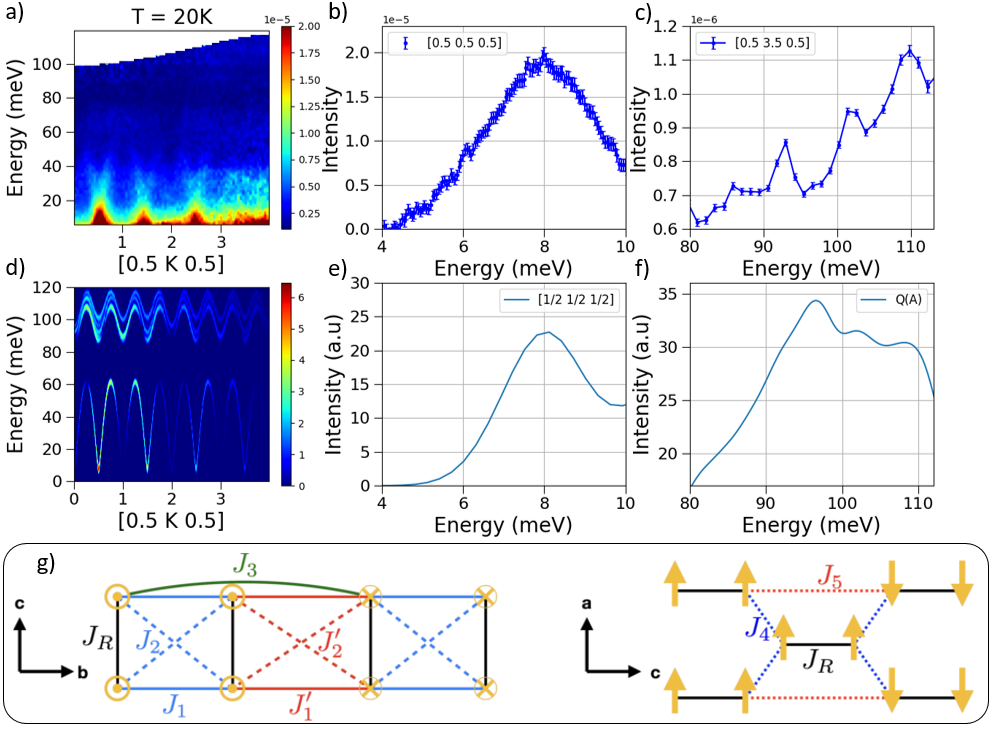}
    \caption{
    a) Color map showing the dispersion along $(1/2, K, 1/2)$ at 20~K ($E_i$ = 125~ meV). Neutron intensity is in logscale. 
    b) Experimental energy cut at $Q=(1/2, 1/2, 1/2)$ taken at 20~K ($E_i$ = 25~meV). The elastic line was removed (see \cite{SI})
    c) Experimental energy cut at $Q=(1/2, 7/2, 1/2)$ taken at 20~K ($E_i$ = 125~meV). The integration range in Q-space was adapted to improve visibility ($\Delta$h=0.5, $\Delta$k=0.5, $\Delta$l=10). 
    d) Simulated color map of $S_{SWT}(Q,\omega)$ along $(1/2, K, 1/2)$. Intensity is in logscale. 
    e) and f) Simulation of the energy cut at $Q=(1/2, 1/2, 1/2)$. 
    g) Schematic sketch of the magnetic structure and the different exchange interactions used in the simulation. Left : projection in the (b,c) plane. Right : projection in the (a,c) plane. Crystallographic axis correspond to the $Pnma$ space group setting. }
    \label{INS map}
\end{figure*}

Single crystals were grown following the method described in Ref.~\cite{Zheng2020}, and fully characterized \cite{SI}. The INS experiments were carried out on the wide angular-range time-of-flight chopper spectrometer (ARCS) at the Spallation Neutron Source (SNS), ORNL, US. The measurements were conducted at 20~K with incident beam energies $E_i$ = 25 and 125~meV yielding an energy resolution of about 1 and 6~meV respectively. Slices were taken from the full $(Q,\omega)$ dataset to produce maps as a function of energy transfer $\omega$ and wavevector $Q$ along the 3 main crystallographic directions $a, b$ and $c$. Only dispersion along the $b$ direction, corresponding to the ladder, was visible, as expected for a quasi uni-dimensional system \cite{SI}. This allowed to integrate along the other directions to improve statistics. The measured dispersion obtained with $E_i$=125~meV at 20 K along $(1/2,K,1/2)$ is shown in Fig. \ref{INS map}a. An acoustic mode is clearly visible, dispersing up to 60~meV. The data obtained with $E_i$=25~meV allows a closer look at the low energy part. By subtracting the elastic part \cite{SI}, a gap is evidenced in the energy cut at $Q=(1/2, 1/2, 1/2)$ as shown in Fig.~\ref{INS map}b. The gap value around 6~meV is in good agreement with the one inferred from powder sample data \cite{Mourigal2015}. At higher energy, three dispersion-less optical modes are observed. In order to make them more visible, an integration along H ($\Delta$H=0.5), K ($\Delta$K=0.5) and L ($\Delta$L=10) was performed, yielding the energy dependence displayed in Fig.~\ref{INS map}c. They appear at 92(2), 101(2) and 110(2)~meV. Only two of these modes were resolved in the powder experiment \cite{Mourigal2015}, providing a strong additional constraint on the subsequent modeling.

\begin{figure}[!htpb]
    \includegraphics[width=1
    \linewidth,angle=0]{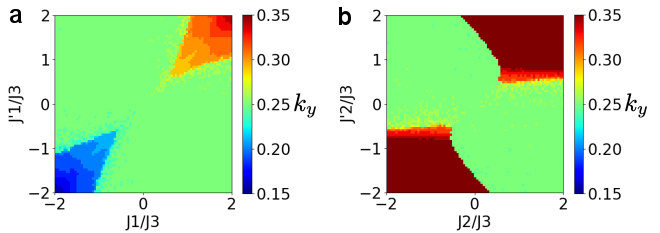}
    \caption {Phase diagram obtained based on the propagation vector determined by Metropolis-Hasting simulations of a single ladder as a function of different couplings. The colorscale indicates the value of $k_y$ along the ladder. In \textbf{a}, the calculations are performed for  $J_2$=$J'_2$=$J_R$=0, while in \textbf{b}, $J_1$, $J'_1$ and $J_R$ values are given in Tab. \ref{tab:J}.}
    \label{MC}
\end{figure}

To model the spin dynamics, 
an exhaustive survey of the structures stabilized by different sets of couplings is necessary. To this end, classical Monte-Carlo was used to determine the phase diagram of a single ladder (see details in \cite{SI}). To describe the system, we used the following Heisenberg model: 
\begin{equation}
\centering
    H_{mag} = \sum_{i<j} J_{ij}\textbf{S}_{i} . \textbf{S}_{j} + \frac{3}{2}D \sum_{i} (\textbf{S}_{i} . \textbf{n}_{i})^2
\label{Heisenberg}
\end{equation}
the first sum runs over iron magnetic ions, $\textbf{S}_{i}$ denote the spin at site $i$ and $J_{ij}$ are exchange integrals between spins at different sites. As sketched in Fig.~\ref{INS map}g, \jun\ and \Jpun\ correspond to nearest neighbor interactions, and \Jtrois\ to next nearest neighbor interactions within a given leg of the ladder. These couplings may lead to frustration in a purely antiferromagnetic model. $J_R$, \Jdeux\ and \Jpdeux\ couple nearest and next nearest neighbor spins respectively in adjacent legs. The second sum describes the magnetic anisotropy : $3D/2$ is the single ion anisotropy and $\textbf{n}_{i}$ defines the local anisotropy axis. 
The results of our calculations are presented in Fig.~\ref{MC}, with maps of the propagation vector value as a function of the various couplings in units of \Jtrois. The unit of the propagation vector is defined by a lattice consisting in a single magnetic site per unit cell. This means that the purely antiferromagnetic order corresponds to $k_y$=0.5 (doubling of the unit cell), and the block order, with 4 spins per unit cell, corresponds to $k_y$=0.25. As can be intuitively expected, provided $J_2=J'_2=J_R$=0, a fully ferromagnetic order with $k_y$=0 is stabilized if $J_1$ and $J'_1$ are ferromagnetic and tend to infinity ($J_1, J'_1 \rightarrow -\infty$). In contrast, a classical antiferromagnetic order with $k_y$=0.5 is stabilized if these parameters are antiferromagnetic ($J_1, J'_1\rightarrow +\infty$), as presented in Fig.~\ref{MC}a. The large green surfaces correspond to the $k_y$=0.25 propagation vector, hence to the up-up-down-down pattern typical of the block order. As expected, this peculiar ordering becomes the ground state as soon as \jun\ and \Jpun\ depart from each other. The symmetry with respect to the diagonal $J_1=J'_1$ is due to the symmetric role of $J_1$ and $J'_1$. The block order is eventually obtained taking into account \Jpdeux\ and \Jdeux, which couple the two legs of the ladder. The map presented in Fig.~\ref{MC}b shows the propagation vector along the ladder for a set of \Jpun\ and \jun\ which stabilizes the up-up-down-down sequence. The values used for \Jpun\, \jun\ and $J_R$ are given in Tab. \ref{tab:J}. Interestingly, the block order remains stable provided that \Jpdeux\ or \Jdeux\ is antiferromagnetic. This indeed contributes to the stabilization of identical up-up-down-down patterns in adjacent legs.

Having reduced the possible range for relative exchange interaction strengths stabilizing the ground state, the spin dynamics was simulated using linear spin wave theory \cite{petit2011numerical}. Each coupling constant $J_i$ was varied independently from the others within the appropriate region of the Monte-Carlo phase diagram to disentangle their role in shaping the dispersion. First, \jun, $J_R$ and \Jdeux\ mainly affect the optical modes bandwidth and energies. \Jtrois, \Jpun\ and \Jpdeux\ affect both acoustic and optical modes. Finally, \Jquatre, \Jcinq\ and the local anisotropy have both an effect on the low energy gap of the acoustic mode. The local anisotropy axis orientation $\textbf{n}_{i}$ has been chosen to reproduce the umbrella-like magnetic ground state \cite{Zheng2022}, with especially a non-zero component along both the $b$ and $c$ directions. The exchange couplings that best fit the experiment are listed in Table \ref{tab:J}. The simulated dispersion is represented Fig.~\ref{INS map}d. As can be seen, the acoustic branches are well reproduced and the 3 optical modes are present. Using the same values, we simulated the energy cut at $Q=(1/2,1/2,1/2)$, displayed in Fig.~\ref{INS map}e, and showing a good agreement with the experimental cut displayed in Fig.~\ref{INS map}b. A powder average has been performed for the high energy branches to compare with the large q-integration of the experimental cut shown in Fig.~\ref{INS map}c. As can be seen in Fig.~\ref{INS map}f, the three experimental modes are reproduced. 

\begin{table}[!htpb]
    \centering
    \caption{Values of exchange interactions (in meV) determined using a purely antiferromagnetic effective Heisenberg model}
    \begin{tabular}{cccccc}
    \hline \hline 
    $J_1$ & 5.5 & $J_R$ & 1.3  & $J_4$ & 0.15\\
    $J'_1$ & 19.2 & $J'_2$ & 18.4 & $J_5$ &  2.0  \\
    $J_3$ & 15.3 & $J_2$ & 6.3  & $D$ & -3.9 \\
     \hline \hline 
    \end{tabular}
    \label{tab:J}
\end{table}

In the present model, the block order is stabilized by strong \Jpun, \Jpdeux\ and \Jtrois. The other couplings are weaker and frustrated. This is at variance with the model proposed in \cite{Mourigal2015}, which assumes the existence of alternating ferromagnetic and antiferromagnetic couplings along each leg of the ladder. Note that single ion anisotropy is also relatively strong in our model. This is at the origin of the strong gap at the magnetic zone center $q=(1/2, 1/2, 1/2)$. Based on our previous analysis of the influence of each coupling, these behaviors lead us to believe that optical modes are related to intra-blocks and frustrated couplings, while the acoustic mode is associated with inter-block couplings and strong anisotropy. It is interesting to note that the energy of this gap nearly coincides with the low energy phonons observed in infrared spectroscopy around 60 cm$^{-1}$ \cite{Weseloh2022}. This particular polar phonon mode seems to be affected by the magnetic ordering below $T_N$, suggesting a coupling with the magnetic excitations. Such coupling between phonon and magnon could indicate the presence of an electromagnon, the famous excitation bearing electroactive and magnetic character \cite{Petit2013, Vaunat2021}.

The new set of exchange parameters determined in the present study presents numerous advantages with respect to the previous proposed model \cite{Mourigal2015}. First, all interactions being antiferromagnetic, the model is compatible with the Goodenough-Kanamori rule and the bonds angle listed in Tab. \ref{tab:expph}. There is also no need to resort to ferromagnetic interaction, an artificial solution that remains rather difficult to justify physically. Second, the new set of parameters describes the 3 modes observed at high energy, and that were not resolved in the previous experiment performed on powder \cite{Mourigal2015}.
The strong amplitude of the nearest-neighbor coupling \Jtrois\ could be partially explained by the chemistry of the chalcogenide. Selenium orbitals are known to be particularly delocalized in space, likely responsible for this large second-neighbor interaction. In the same vein, Sulfur orbitals are expected to be less delocalized, resulting in a weaker second-neighbor interaction in BaFe2S3. This explains why the nearest-neighbor frustrated coupling is lifted in this compound, leading to an antiferromagnetic stripe-like order.
Third, the magnetic transition observed under pressure in \cite{zheng2022universal} finds a natural explanation in the present model. 
Indeed, as shown by Monte Carlo calculations, the magnetic propagation wavevector is quite sensitive to the \jun/\Jtrois\ ratio, with the stripe-like phase being found when \Jtrois\ decreases. The intuitive decrease of \Jtrois\ upon pressure is further supported by first-principle calculations performed on the parent compounds of \bfs, namely FeSe and FeTe \cite{Glasbrenner2015}. The authors claim that under pressure, the ground state of these systems transitions to an antiferromagnetic stripe order, as is the case for \bfs\ at 4 GPa.

However, detailed calculations of the exchange couplings would be helpful : their evolution as a function of pressure is related to the structure in a non-trivial way. Fourth, the model detailed in this work explains the stripe antiferromagnetic order observed in all other members of the series with similar structure \cite{Caron2012,hawai2015,Takahashi2015}. Finally, the frustration inherent to this model provides a key ingredient to elucidate the magneto-electric coupling in this system. Indeed, it has been shown that this compound is multiferroic \cite{Zheng2020, Du2020, Weseloh2022}, and exhibits a significant structural change across the N\'eel temperature, with an increased deviation from the non-polar average space group \cite{Zheng2020}. While the magnetic structure chirality reported recently suggests that the inverse Dzyaloshinski-Moriya interaction could be a relevant ingredient at play \cite{Zheng2022}, the presence of frustration suggests that an exchange-striction mechanism could also be at work. This reinforces the picture of a strong magneto-elastic coupling in this systems as suggested in ref \cite{Weseloh2022}.

In conclusion, on the basis of single crystal inelastic neutron scattering experiments, is infered an antiferromagnetic Heinsenberg model, providing a natural explanation for numerous properties observed in this quasi unidimensional iron based multiferroic compound. This new model can be easily extended to other members of the family by tuning the parameters to describe the underlying magnetic orders. It would be interesting to revisit the already published spin dynamics data \cite{Wang2016,Wang2017} in the light of this new model and use it also for future studies of magnetic excitations. This result finally provides a new starting point to study the remarkable properties of these systems, among which multiferroicity and superconductivity.



Aknowledgment : 
The research performed at ARCS used resources at Spallation Neutron Source and was supported by the U.S. Department of Energy, Office of Science, User Facilities Division, and the National Quantum Information Science Research Centers, Quantum Science Center, operated by Oak Ridge National Laboratory. This work was financially supported by ANR COCOM 20-CE30-0029 and FRAGMENT 19-CE30-0040. We acknowledge the MORPHEUS platform at the Laboratoire de Physique des Solides for sample alignment. 
\bibliography{BFS.bib}
\end{document}



\title{Purely antiferromagnetic frustrated Heisenberg model in spin ladder compound BaFe$_2$Se$_3$ : Supplemental information}

\author{A. Roll}
\affiliation{Universit\'e Paris-Saclay, CNRS, Laboratoire de Physique des Solides, 91405, Orsay, France.}
\affiliation{Laboratoire L\'eon Brillouin, CEA, CNRS, Universit\'e Paris-Saclay, 91191, Gif sur Yvette, France}

\author {S. Petit}
\affiliation{Laboratoire L\'eon Brillouin, CEA, CNRS, Universit\'e Paris-Saclay, 91191, Gif sur Yvette, France}

\author{A. Forget}
\affiliation{Universit\'e Paris-Saclay, CEA, CNRS, SPEC, 91191, Gif-sur-Yvette, France.}

\author{D. Colson}
\affiliation{Universit\'e Paris-Saclay, CEA, CNRS, SPEC, 91191, Gif-sur-Yvette, France.}

\author{A. Banerjee}
\affiliation{Neutron Scattering Division, Oak Ridge National Laboratory, Oak Ridge, Tennessee 37831, USA}
\affiliation{Department of Physics and Astronomy, Purdue University, West Lafayette, IN 47906, USA}

\author{P. Foury-Leleykian}
\affiliation{Universit\'e Paris-Saclay, CNRS, Laboratoire de Physique des Solides, 91405, Orsay, France.}

\author{V. Bal\'edent}
\affiliation{Universit\'e Paris-Saclay, CNRS, Laboratoire de Physique des Solides, 91405, Orsay, France.}
\email [Corresponding author: ] {victor.baledent@universite-paris-saclay.fr}

\date{\today}


\maketitle


In this document, we give additional information relative to the sample quality, as well as further details about the anisotropy gap and spin dynamics.
    
\section{Sample mosaic}

\begin{figure}[!ht]
    \centering
    \includegraphics[width=1\linewidth, angle=0]{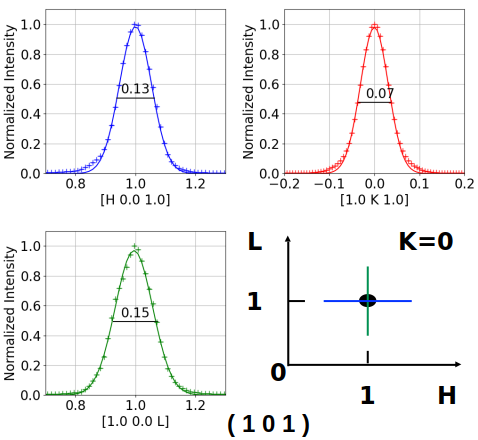}
    \caption{Cuts along H K and L of the elastic (E=0meV) nuclear Bragg peak  $(1,0,1)$. Errorbars are contained in the symbol size. Lines are Gaussian fits.}
    \label{gau-nucl}
\end{figure}

To determine the sample mosaic, cuts along H, K and L directions were extracted around the $(1,0,1)$ Bragg reflection. Integration parameters for these cuts are [0.9:1.1], [-0.1:0.1], [-0.1:0.1] and [-2:2] along H, K, L and energy respectively. Fitting a Gaussian function to the data, as shown in Fig. \ref{gau-nucl}, we determined the widths to be $\Delta H=$~0.13, $\Delta K=$~0.07, $\Delta L=$~0.14 in reciprocal lattice unit. This corresponds to correlation length of 14 units cells along $b$ (the ladder direction) and 7 units cells along $a$ and $c$. These correlation lengths are smaller than the one observed in ref. \cite{Zheng2020} by X-ray diffraction, and can be easily explained by the much smaller size of the sample used : typically 30x30x100$\mu$m$^3$ for X-ray diffraction and around 0.5x1x1 cm$^3$ here.

\section{Magnetic ordering}

One can see that the magnetic order develops at quite short range : 5,
9 and 2 unit cells along H, K and L respectively . This is probably due to impuritites in the 1D magnetic structure.

\begin{figure}[!ht]
    \centering
    \includegraphics[width=1\linewidth, angle=0]{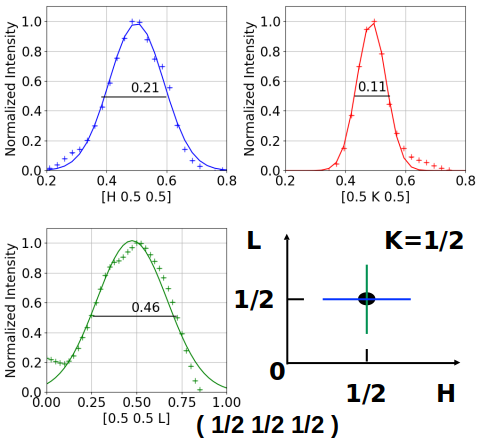}
    \caption{Cuts along H K and L of the elastic (E=0meV) magnetic Bragg peak  $(1/2,1/2,1/2)$. Errorbars are contained in the symbol size. Lines are Gaussian fits. }
    \label{gau-mag}
\end{figure}

To characterize the magnetic order, similar cuts along H, K and L directions were extracted around the $(1/2, 1/2, 1/2)$ magnetic Bragg reflection, see Fig \ref{gau-mag}. As one can see, the magnetic order develops at shorter range than the atomic structure with magnetic correlation lengths of 5, 9 and 2 unit cells along H, K and L respectively. This is probably due to impuritites in the 1D magnetic structure.\\

\begin{figure}
    \centering
    \includegraphics[width=1\linewidth, angle=0]{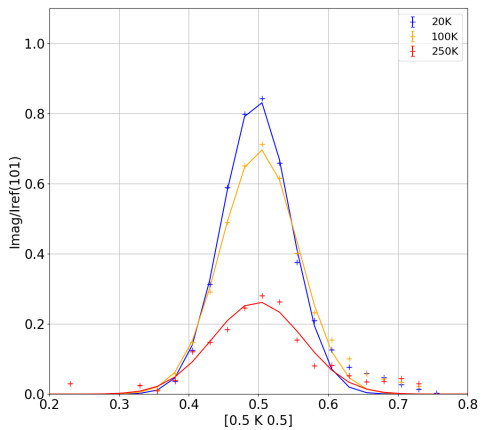}
    \caption{Cuts along K of the elastic (E=0meV) magnetic Bragg peak  $(1/2,1/2,1/2)$ for different temperature T=20, 100 and 250 K. Errorbars are contained in the symbol size. Lines are Gaussian fits.}
    \label{fig:temp}
\end{figure}

In order to check the sample quality, we fitted the intensity of the $(1/2, 1/2, 1/2)$ for the  three temperatures available, namely 20 K, 100 K and 250 K. Indeed, the Neel temperature is in direct relationship with the sample quality.  As can bee seen in Fig. \ref{fig:temp}, from 20K, to 100 K, the
intensity is nearly constant, and is strongly reduced at 250 K, suggesting
a T$_N$ close to 200 K as reported in other batches using the very same synthesis conditions (see Fig. 4 in supplementary information of ref. \cite{Zheng2022}).
At 250 K, the peak width is sligthly larger with a reduced but non negligable intensity, suggesting a fluctuating magnetic order, compatible with a Neel temperature around 200 K. 

As can bee seen in Fig. 3, from 20K, to 100 K, the
intensity is nearly constant, and is strongly reduced AT 200 K, suggesting
a TN close to 200 K as reported in other batches
using the very same synthesis conditions (see FIG 4 in supplementary
information of ref. [1]).

\section{Anisotropy gap determination}

To evidence the anisotropy gap, we removed the elastic contribution to the energy cut at $(1/2, 1/2, 1/2)$ position from the raw data. The raw data together with the fit is presented in Fig. \ref{fig:lor}. The background function used in the fit is a combination of a lorentzian fit for the elastic intensity, as well as a background noise with a flat gaussian. The background-substracted intensity is presented in the main manuscript to be compared with simulations.

\begin{figure}
    \centering
    \includegraphics[width=1\linewidth, angle=0]{Images/[0.5 0.5 0.5]_gap.png}
    \caption{Intensity at q=$(1/2, 1/2, 1/2)$ as function of energy (points) with the fitted background (line).}
    \label{fig:lor}
\end{figure}

\section{Spin wave dispersion along H and L}

\begin{figure}[!htpb]
    \includegraphics[width=1\linewidth, angle=0]{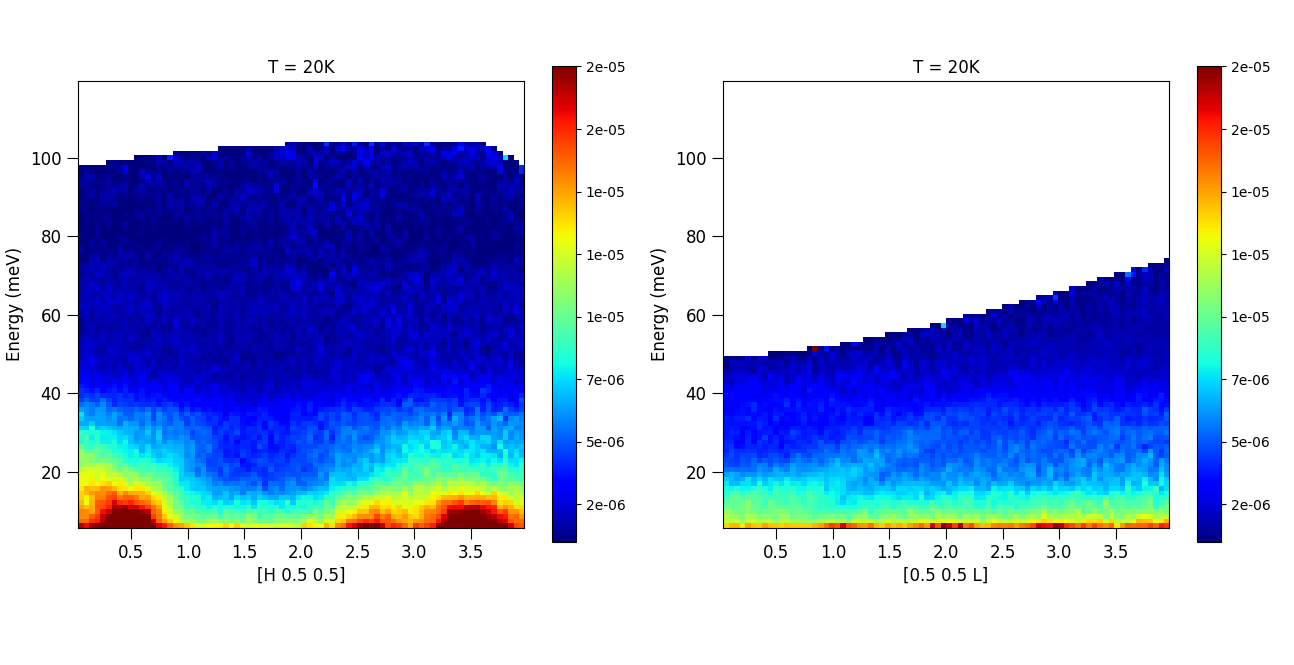}
    \caption{Color map along $(H, 1/2, 1/2)$ (left) and $(1/2, 1/2, L)$ (right) of the dispersion  at 20~K ($E_i$ = 125~ meV), with integration range of $\Delta K$=0.5 $\Delta H$=0.5 and $\Delta L$=10 }
    \label{fig:dispersionsHL}
\end{figure}

We extracted the dispersion along H and L around the $(1/2,1/2,1/2)$ magnetic Bragg peak for $E_i$=125meV. Using the integration parameters $\Delta K$=0.5 for both maps, and $\Delta H$=0.5 and $\Delta L$=10, the results are presented in Fig. \ref{fig:dispersionsHL}. As we can see, no dispersion is visible for both directions, as expected from the short range correlations extracted from the elastic peak along a and c axis of 5 and 2 unit cells respectively. 


\section{Monte-Carlo simulations}

\begin{figure}
    \centering
    \includegraphics[width=1\linewidth, angle=0]{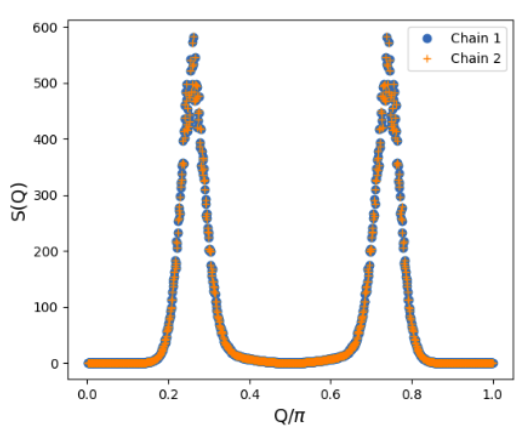}
    \caption{Magnetic structure factor S(Q) for both legs of a ladder with the couplings taken in the main article (see Tab. II).}
    \label{fig:Sq}
\end{figure}

 Classical Monte-Carlo, with both Metropolis-Hasting algorithm and adaptive method described in \cite{alzate2019optimal}, was used to determine the phase diagram of a single ladder. In the calculations, a ladder consisting of 2 legs of 100 spins coupled together was considered, and the simulations were performed with 10 000 Monte Carlo Steps (MCS) to ensure convergence. In these simulations, no simulated annealing was performed and the temperature was set to 0.01 K. The magnetic structure factor $S(Q)$ was then computed, averaging over the ground state configurations sampled by the Metropolis-Hasting algorithm. The propagation vector of the structure was determined based upon the $Q$-value which maximizes $S(Q)$. The Fig.\ref{fig:Sq} shows the calculations of S(Q) for the exchanges couplings of our model: we observe 2 distincts peaks, which are actually the same, because of the periodicity and symmetry of the Brillouin zone. The first peak is centered around k = 0.25, well defined in the main article as the propagation vector of the expected structure.

\bibliography{BFS.bib}